\documentclass[11pt]{article}
\usepackage{moriond,epsfig}

\def\Journal#1#2#3#4{{#1} {\bf #2}, #3 (#4)}

\def\NPB{{\em Nucl. Phys.} B}
\def\PLB{{\em Phys. Lett.}  B}

\def\PRD{{\em Phys. Rev.} D}

\def\JPC{{\em Eur. Phys. J.} C}
\def\JPG{{\em J. Phys.} G}

\def\be{\begin{equation}}
\def\ee{\end{equation}}
\def\bea{\begin{eqnarray}}
\def\eea{\end{eqnarray}}

\begin{document}
\vspace*{1.5cm}
\title{Jet Physics and Event Shape Studies at HERA}

\author{ E. Rodrigues }

\address{H. H. Wills Physics Laboratory, Tyndall Avenue,\\
Bristol BS8 1TL, England\\[0.2cm]
(on behalf of the ZEUS and H1 Collaborations)}

\maketitle\abstracts{
A review is given of the latest results on jet production and studies of
event shape variables in deep inelastic scattering at HERA.
Jet cross sections studies for inclusive jet and dijet events are presented and
compared to next-to-leading order QCD calculations. Extraction of the strong coupling
constant $\alpha_s$ is discussed.}

\section{Introduction}
At HERA, studies of the hadronic final state of electron-proton deep inelastic
scattering (DIS) $ep \to eX$ allow to test QCD over a large range of $Q^2$,
$Q$ being the virtuality of the probing boson.
We will here focus on the latest measurements of jet production cross sections
and event shape variables. The sensitivity of the measured observables to the
value of $\alpha_s$ will be exploited to determine its value.

In the Breit frame, the production of high transverse energy events can be
related to the gluons emitted in the hard QCD process. Most of the
analyses were carried out in this frame.

\vspace*{-0.3cm}
\unitlength1.0cm
\begin{figure}[h]
\begin{center}
\begin{picture}(13.0,7.5)
\put(0,-0.5){\scalebox{0.35}{\epsfig{figure=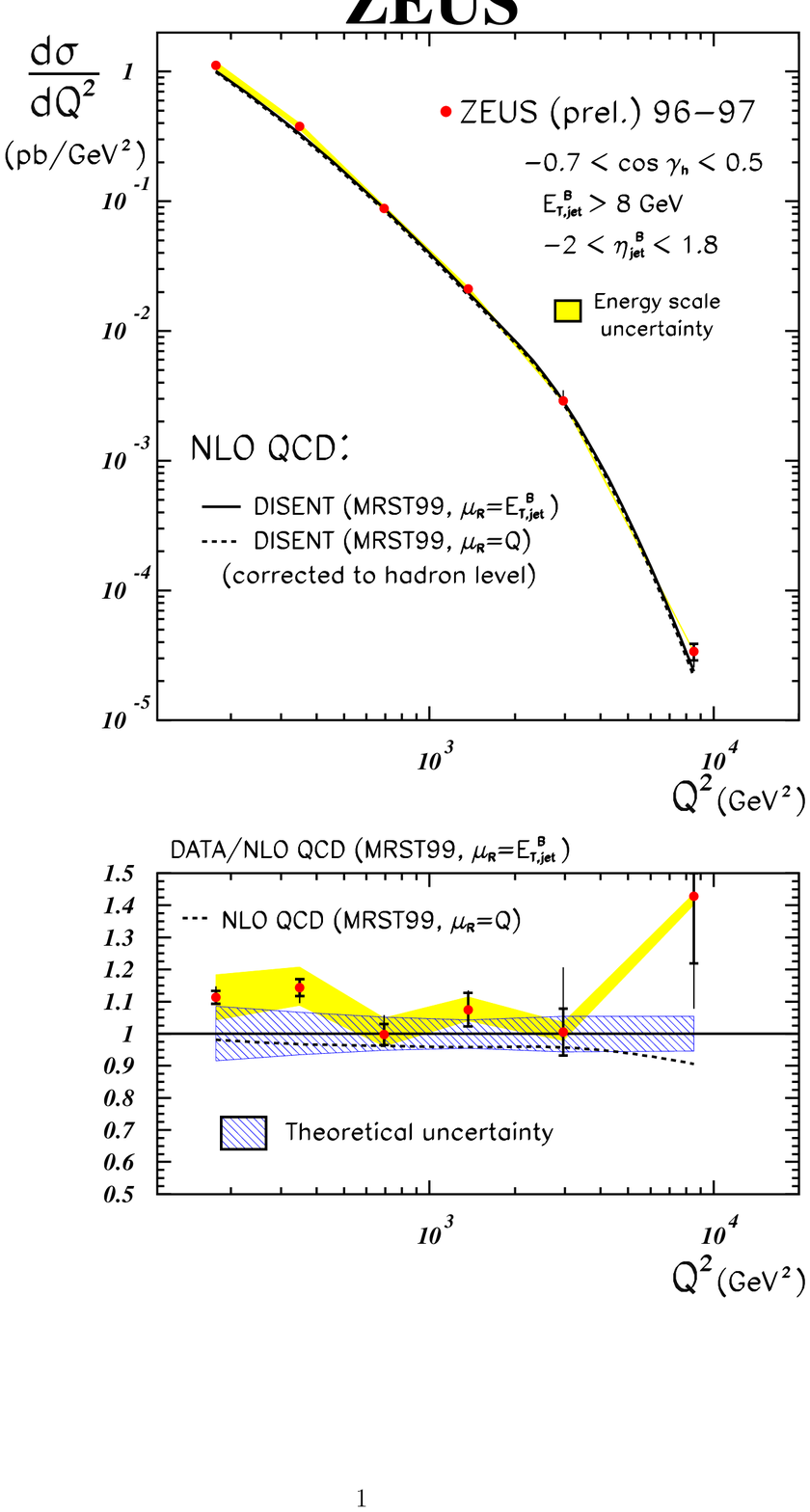}}}
\put(6.0,0.){\scalebox{0.35}{\epsfig{figure=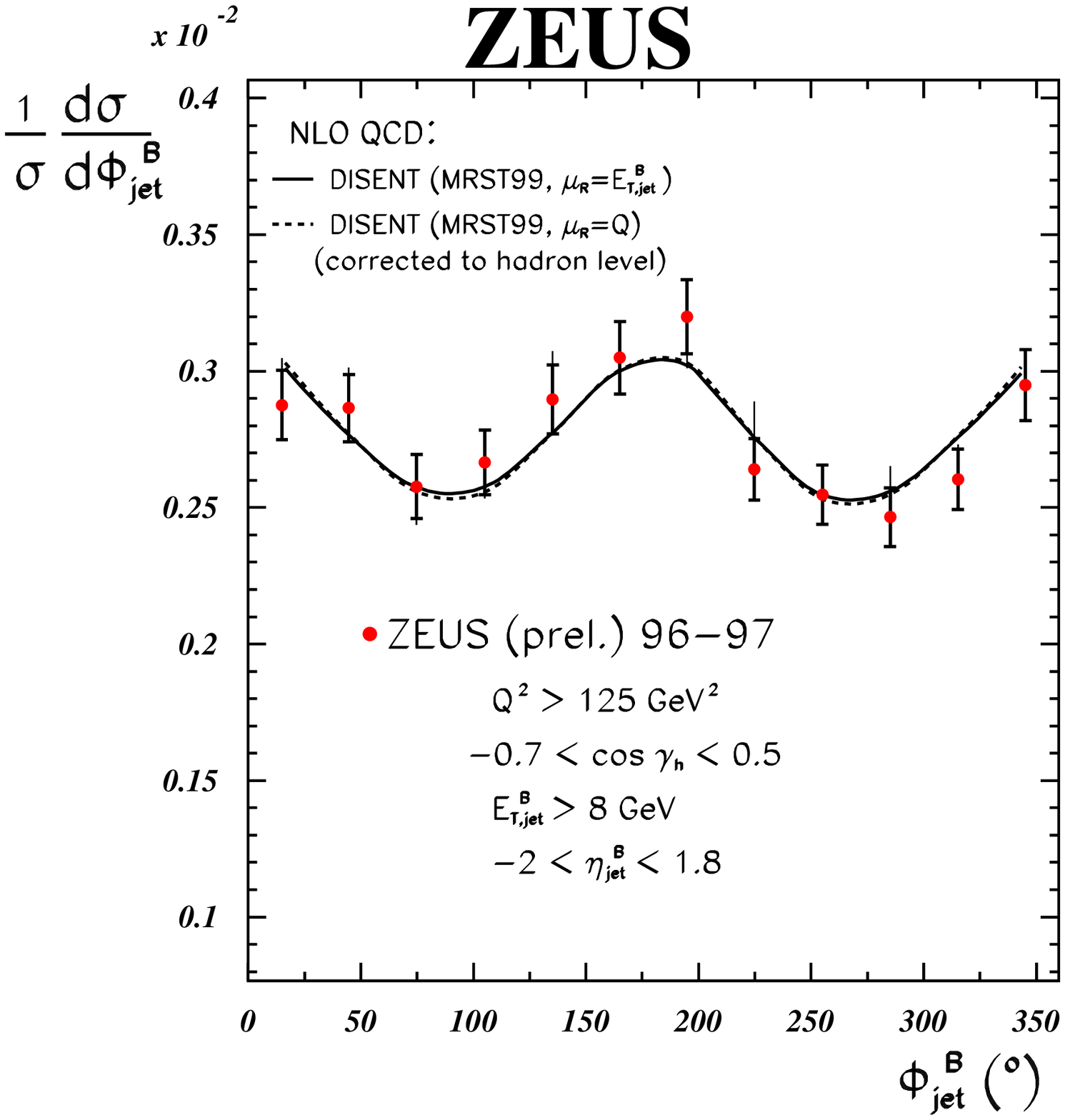}}}
\put(-0.5,3.5){ a) }
\put(6.,3.5){ b) }
\end{picture}
\end{center}
\caption{Inclusive cross section measurements in the Breit frame as a function of
        (a) $Q^2$ and (b) $\phi^B_{jet}$, where $\phi^B_{jet}$ is the jet's azimuthal
        angle as measured with respect to the lepton scattering plane.
        \label{fig:zeus_incl}}
\end{figure}

\vspace*{-0.3cm}
\section{Jet production in DIS}
Studies of inclusive jet production provide a powerful means of investigating
QCD: the DIS inclusive jet cross section definition does not present the
infrared-sensitivity problems related to the jet selection, as in the case
of dijet production~\cite{bib:etjetcuts}.
And furthermore, the inclusive jet cross section is very
sensitive to the value of the strong coupling constant $\alpha_s$, hence
providing a powerful method of extraction of its value.

ZEUS performed an analysis of inclusive jet production in DIS in the Breit
frame~\cite{bib:zeus_incl}. Jets were found in the Breit frame with the inclusive
$k_T$ clustering algorithm~\cite{bib:ktincl}, and only jet cuts in the Breit frame
were applied. Figure~\ref{fig:zeus_incl}(a) shows the inclusive jet cross section as a
function of $Q^2$. The set of selection cuts is displayed in figure~\ref{fig:zeus_incl}.

Next-to-leading order (NLO) QCD calculations are able to describe the data over the
5 orders of magnitude within $10-15\%$.
The value of $\alpha_s$ as determined from a comparison of the data to the NLO
QCD calculations in the high $Q^2$ region ($Q^2 > 500 \textrm{ GeV}^2$)
is~\cite{bib:zeus_incl}
\be
\alpha_s (M_Z)
   = 0.1190 \pm 0.0017 (\mathrm{stat})
     {\textstyle{+ 0.0049 \atop - 0.0023}} (\mathrm{syst})
     {\textstyle{+ 0.0026 \atop - 0.0026}} (\mathrm{th}).
\ee
For comparison, H1 has found a value of~\cite{bib:hi_alphas}
\be
\alpha_s (M_Z)
   = 0.1186 \pm 0.0030 (\mathrm{exp})
     {\textstyle{+ 0.0039 \atop - 0.0045}} (\mathrm{th})
     {\textstyle{+ 0.0033 \atop - 0.0023}} (\mathrm{pdf}).
\ee
in similar studies of inclusive jet production.
Both results are in excellent agreement with the world average~\cite{bib:alphas_worldavg}
$\alpha_s (M_Z) = 0.1184 \pm 0.0031$.

The same ZEUS analysis was also used to measure for the first time the
azimuthal angle distribution of jets in the Breit frame, with respect to the
lepton scattering plane (figure~\ref{fig:zeus_incl}(b)).
The shape of the distribution, of the form
\be
\frac{d\sigma}{d\phi^B_{jet}} = A + C \cos2\phi^B_{jet}\quad,
\ee
is in very good agreement with NLO QCD, and provides a detailed test of the QCD matrix
elements of the hard process.

\vspace*{-0.2cm}
\unitlength1.0cm
\begin{figure}[h]
\begin{minipage}[b]{8.0cm}
Two possible hard energy scales are natural for the renormalisation scale
$\mu_R$ in NLO QCD calculations: $Q$ and the transverse energy, $E_T$,
of the event. H1~\cite{bib:h1_dijetrate} investigated the effect of this
choice in dijet production at medium $Q^2$. The study was done in the
photon-proton centre-of-mass frame in the kinematic region defined by
$10^{-4}\!<\!x\!<\!10^{-2}$ and $5\!<\!Q^2\!<\!100 \mbox{\textrm{ GeV}}^2$.
In this region of relatively low $x$ and $Q^2$, where higher order corrections
are expected to be larger, the choice of $\mu_R$ is still a matter of
discussion; and $Q$ is of the same order as the jets transverse energies.

The jets obtained with the $k_T$-cluster algorithm were selected with
$E_{T}^{jet,1} > 7 \mbox{\textrm{ GeV}}$ and
$E_{T}^{jet,2} > 5  \mbox{\textrm{ GeV}}$
in the pseudorapidity range $-1\!<\!\eta^{jet}_{lab}\!<\!2$.
Figure~\ref{fig:h1_dijetrate} shows the dijet rate as a function of Bjorken $x$
in different regions of $Q^2$.
The NLO calculations with $\mu_R^2 = Q^2$ are able to describe the data but
present large scale uncertainties, whereas the choice
$\mu_R^2 = Q^2 +  \overline E_{T}^{jet}$ 
results in smaller scale uncertainties but fails to decribe the distributions.
\end{minipage}
\hfill
\begin{minipage}[b]{7.0cm}
\begin{picture}(7.0,6.0)
\scalebox{0.4}{\psfig{figure=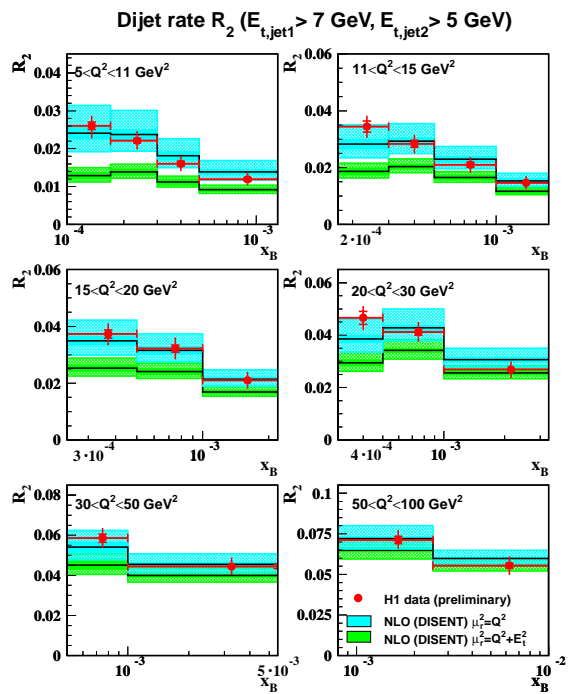}}
\end{picture}\par
\caption{Dijet rate as a function of Bjorken $x$ in different regions of $Q^2$.
         Data is compared with NLO QCD calculations with 2 different choices of the
         renormalisation scale $\mu_R$. \label{fig:h1_dijetrate}}
\end{minipage}
\end{figure}
\vspace*{-0.3cm}
At higher $Q^2$ the particular choice of $\mu_R$ has a smaller impact on the
perturbative QCD (pQCD) predictions, which describe reasonably well the dijet
data~\cite{bib:dijets}. Also the theoretical uncertainties become smaller.
But these jet analyses usually require a large inter-jet separation: either
the jets have a large relative transverse energy or they are selected with a large
transverse momentum, to avoid the effects of multi-parton emissions becoming
significant. About $10\%$ of the DIS sample is then classified as
dijet events in these ``standard'' jet analyses performed at HERA.

H1 has investigated the minimum jet separation necessary for the NLO calculation
to give an accurate description of dijet production~\cite{bib:h1_jetsep}.
The study was undertaken in a region of relatively high $Q^2$:
$150 < Q^2 < 35000 \mbox{\textrm{ GeV}}^2$ and $0.1 < y < 0.7$\,.
The jets were reconstructed in the laboratory frame with the modified Durham
algorithm~\cite{bib:Durham}. Dijet events were analysed by means of the
variable
$y_2 = \frac{\min{k_{Ti,j}^2}}{W^2}$
($W$ is the invariant mass of the particles clustered),
where
$k_{Ti,j}^2 = 2 \min [E_i^2,E_j^2](1-\cos\theta_{ij})$
is the relative measure of the separation between jets $i$ and $j$, and
$\theta_{ij}$ the angle between them.
\vspace*{-0.2cm}
\unitlength1.0cm
\begin{figure}[h]
\begin{minipage}[b]{7.5cm}
Figure~\ref{fig:h1_dijetsep} shows the measured $y_2$ distribution normalised
to the inclusive DIS cross section $\sigma_{DIS}$. NLO QCD predictions are in
good agreement with the data for values of $y_2 > 0.001$,
with about $1/3$ of the events being classified as dijet events.
But it overestimates the data dramatically in the region where the inter-jet separations
are small. Such behaviour was expected since in the small $y_2$ region the difference
between LO and NLO is the largest, and also the renormalisation scale uncertainty
and the hadronisation corrections are large, making the NLO calculations
(fixed order) unreliable.
The RAPGAP LO Monte Carlo program is seen to be in good
agreement with the data over all the $y_2$ range.
\end{minipage}
\hfill
\begin{minipage}[b]{7.5cm}
\begin{picture}(7.5,6.0)
\scalebox{0.4}{\psfig{figure=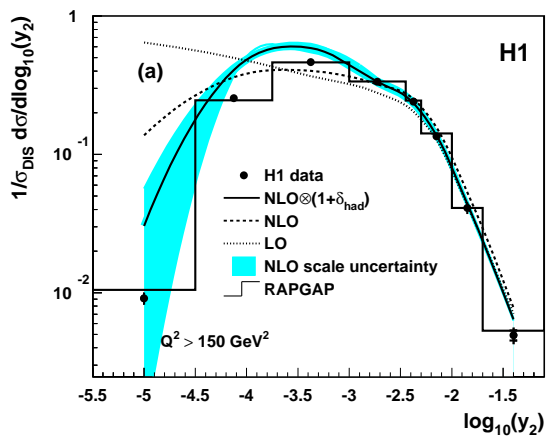}}
\end{picture}\par
\caption{The measured $y_2$ distribution normalised to the inclusive DIS
         cross section. The data were compared with LO and NLO QCD calculations,
         and the RAPGAP Monte Carlo program.
\label{fig:h1_dijetsep}}
\end{minipage}
\end{figure}

\vspace*{-0.4cm}
\noindent
The dijet sample defined by $y_2 > 0.001$ was then further analysed. NLO QCD
calculations were found to describe well the distributions of several relevant jet
variables.

\section{Event shape variables}
Event shape variables are interesting observables that are sensitive to the overall
topology of the event and therefore to higher order QCD radiation.
Within the QCD framework the mean of an event shape $F$ is given by
\be
<F>(Q) = \underbrace{<F>^{pQCD}}_{\mathcal{O}(\alpha_s^2)}
         + \underbrace{<F>^{Pow.Corr.}}_{f(\alpha_s,\overline \alpha_0)} \, , 
\ee
the pQCD part being calculated at present to next-to-leading order. The second term in
the above equation, the ``power correction'', refers to the non-perturbative part,
necessary for the theory to describe the data. This non-perturbative contribution which
accounts for hadronisation corrections is expressed as a power law correction, which
depends on the value of the strong coupling constant $\alpha_s$ and on an empirical
non-perturbative parameter $\overline \alpha_0$~\cite{bib:powcorr}.

The mean values as a function of $Q$ of the thrust, broadening, jet mass and C-parameter
where studied in the Breit frame by both the ZEUS~\cite{bib:zeus_evtshp} and
H1~\cite{bib:h1_evtshp} collaborations. A general good agreement was found between the
ZEUS and H1 results. The simultaneous 2-dimensional fit to all
mean event shapes yielded $\overline \alpha_0 \approx 0.5 \pm 20\%$ and a relatively
spread range of values for $\alpha_s (M_Z)$, suggesting the need for inclusion of higher
order terms in the theoretical calculations.

\begin{figure}[h]
\begin{center}
\begin{picture}(14.0,3.6)
\put(0.,3.7){\scalebox{0.25}{\epsfig{figure=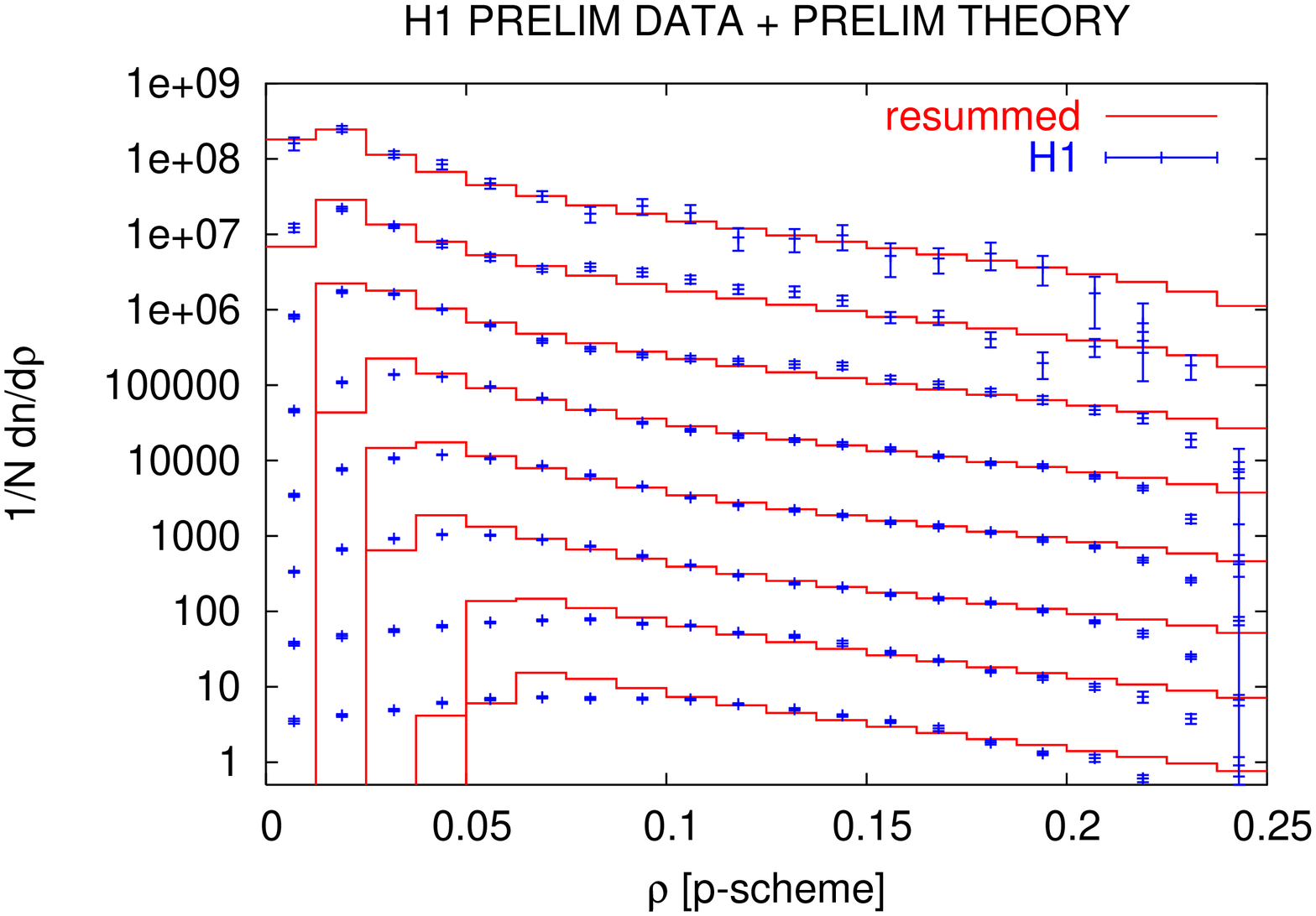,angle=-90}}}
\put(7.0,3.7){\scalebox{0.25}{\epsfig{figure=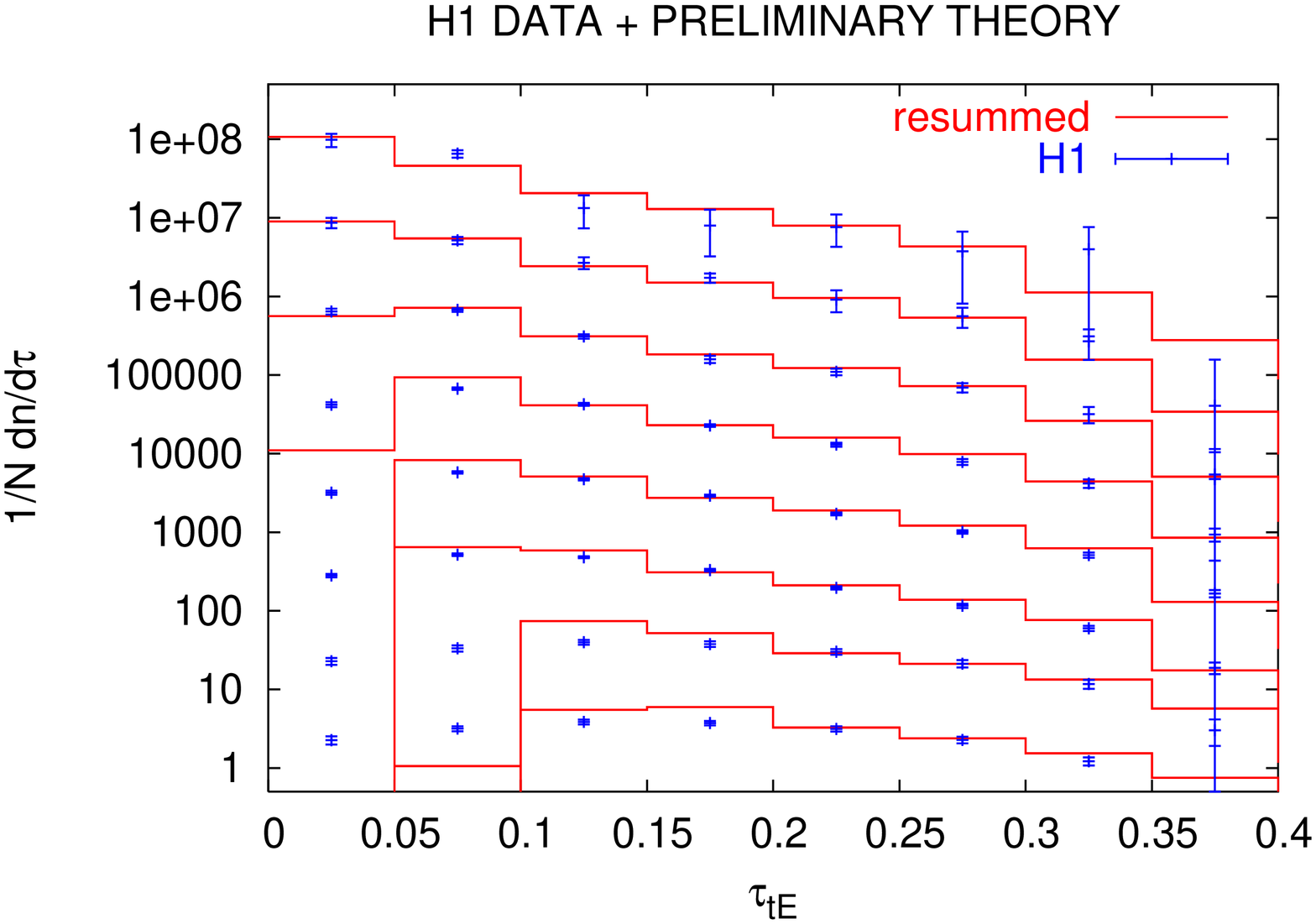,angle=-90}}}
\put(-0.5,2.5){ a) }
\put(6.6,2.5){ b) }
\end{picture}
\end{center}
\caption{Differential distributions for the (a) jet mass $\rho$
        and (b) the thrust $\tau_{tE} = 1 - T$: fits to the
        H1 data including QCD resummed calculations with power corrections.
        \label{fig:h1_evtshp}}
\end{figure}

H1 has also performed fits to the differential distributions of the event
shapes~(see~\cite{bib:h1_evtshp_dist}). The values obtained for $\alpha_s$
and $\overline \alpha_0$ were found to be inconsistent with those obtained
from the fits to the means of the event shapes.
Fits using QCD resummed calculations improve significantly the description
of the differential distributions of the event shapes
(\textit{cf.} figure~\ref{fig:h1_evtshp}~\cite{bib:h1_evtshp_resum}).
Latest theoretical work~\cite{bib:Salam} suggests that a consistent picture
can be obtained when applying the concept of resummed QCD calculations
in a simultaneous fit to all the event shape distributions.

\unitlength1.0cm

\vspace*{-0.5cm}
\section*{Acknowledgements}
I would like to acknowledge the financial support given by the
Funda\c c\~ao para a Ci\^encia e a Tecnologia
and the University of Bristol.

\section*{References}

\end{document}